\documentclass{Interspeech2024}

\usepackage{amsmath,graphicx}
\usepackage[svgnames]{xcolor}

\usepackage{booktabs} 
\usepackage{multirow}
\usepackage{enumitem}
\usepackage{comment}

\usepackage{nicefrac}       

\usepackage{hyperref}       

\interspeechcameraready

\title{TacoLM: GaTed Attention Equipped Codec Language Model are Efficient Zero-Shot Text to Speech Synthesizers}

\name[affiliation={1}]{Yakun}{Song}
\name[affiliation={2}]{Zhuo}{Chen}
\name[affiliation={3}]{Xiaofei}{Wang}
\name[affiliation={1}]{Ziyang}{Ma}
\name[affiliation={1}]{Guanrou}{Yang}
\name[affiliation={1,\dagger}]{Xie}{Chen}

\address{
  $^1$MoE Key Lab of Artificial Intelligence, AI Institute, X-LANCE Lab, \\ Shanghai Jiao Tong University, China\\
  $^2$ByteDance  $^3$Microsoft, One Microsoft Way, Redmond, USA}
\email{\{ereboas, chenxie95\}@sjtu.edu.cn}


\newcommand{\underlinebold}[1]{\underline{\textbf{#1}}}
\newcommand\blfootnote[1]{
  \begingroup
  \renewcommand\thefootnote{}\footnote{#1}
  \addtocounter{footnote}{-1}
  \endgroup
}

\begin{document}

\maketitle

\begin{abstract}
    
    Neural codec language model (LM) has demonstrated strong capability in zero-shot text-to-speech (TTS) synthesis. However, the codec LM often suffers from limitations in inference speed and stability, due to its auto-regressive nature and implicit alignment between text and audio. In this work, to handle these challenges, we introduce a new variant of neural codec LM, namely TacoLM. Specifically, TacoLM introduces a gated attention mechanism to improve the training and inference efficiency and reduce the model size. Meanwhile, an additional gated cross-attention layer is included for each decoder layer, which improves the efficiency and content accuracy of the synthesized speech.
    In the evaluation of the Librispeech corpus, the proposed TacoLM achieves a better word error rate, speaker similarity, and mean opinion score, with 90\% fewer parameters and 5.2 times speed up, compared with VALL-E. 
    Demo and code is available at https://ereboas.github.io/TacoLM/.
\end{abstract}

\blfootnote{$\dagger$ Corresponding author}

\section{Introduction}
\label{sec:intro}

With the development of deep neural networks, text-to-speech (TTS) technology has made significant progress~\cite{glowtts,gradtts,paralleltacotron,speartts}. Among them, zero-shot TTS only requires a short audio sample as prompts to synthesize high-quality speech for any unseen speaker. Zero-shot TTS does not need to be fine-tuned with speech data with respect to new speakers, which is of more practical value but still challenging. Early studies use continuous audio signals as input, relying on an explicit speaker encoder to embed a speaker's timbre, prosody, and speaking style~\cite{scglowtts,yourtts}. Some studies further rely on specifically designed speech disentanglement approaches to extract speaker-agnostic information~\cite{kumar2022zero,generspeech}. However, when the speaker embedding is inaccurate, the model's ability to generalize to zero-shot scenarios decreases dramatically. 

Recent developments in diffusion and language modeling bring breakthroughs to the zero-shot TTS. The former~\cite{NS2,voicebox,fastdiff} leverages the diffusion model~\cite{ho2020ddpm} and its variants~\cite{flowmatching,yang2022diffusion} to estimate the target speech that shares the same distribution as the enrollment, while the latter~\cite{audiolm} usually employs language models on a pre-trained neural codec to extract discrete audio representations and reconstruct high-quality waveforms. Both systems achieve impressive performance in the field of zero-shot speech synthesis. As the neural codec language model doesn't require a separated duration prediction model, which potentially enables a more direct end-to-end optimization, we focus on the improvement of this algorithm family in this work. 
In the domain of neural codec language models, VALL-E~\cite{valle} is a prototypical and highly effective two-stage approach. Specifically, VALL-E takes the phoneme and acoustic tokens as prompts, and leverages an autoregressive (AR) and a non-autoregressive (NAR) language model, to generate coarse and fine-grained acoustic tokens of the unenrolled speaker, respectively. This powerful approach obviates the need for encoding the speaker into embeddings, while allowing for direct extraction of the environment and acoustic information from the audio samples. 

Despite its achievements, the VALL-E still suffers from many drawbacks. One drawback that affects the practical application experience is the slow speed, arising from the use of an AR model to generate coarse-grained acoustic tokens. While the multi-head attention mechanism in the Transformer can model the relationship between pairs of tokens well, it often faces large computational and memory costs. The speed deficiency is especially prominent in the inference process, since the AR model needs to generate outputs in a token-by-token manner. Another limitation of VALL-E is the occasional mismatch between the synthesized speech and text prompts, which can manifest as repetitions, transpositions, or omissions. This is because the model does not align text and speech well. How to realize zero-shot TTS in both an efficient and accurate manner remains an open challenge.

In this study, we address the zero-shot TTS task with our proposed TacoLM (Ga\underlinebold{T}ed \underlinebold{A}ttention Equipped \underlinebold{Co}dec \underlinebold{L}anguage \underlinebold{M}odel), which is a two-stage (AR + NAR) framework, similar to VALL-E.
TacoLM incorporates a MEGA module, as detailed in \cite{mega}, which is based on a single-head gated attention mechanism with an exponential moving average. 
As a result, TacoLM is computationally efficient and requires minimal memory and storage.
In addition, TacoLM employs a gated cross-attention mechanism to exchange information between text and audio, aiming at enhancing the accuracy of the synthesized speech.
To evaluate TacoLM in the zero-shot scenario, we conducted experiments on the LibriSpeech~\cite{librispeech} dataset. The experimental results show that TacoLM is superior to the advanced baseline VALL-E in terms of both objective (WER and speaker similarity) and subjective metrics (CMOS and SMOS). 
Ablation studies further corroborate the individual contributions of TacoLM's components to its overall effectiveness.
Our contributions can be summarized as follows:
\begin{itemize}[itemsep=0.5mm, parsep=1pt, leftmargin=*]
    \item We propose TacoLM, a two-stage zero-shot text-to-speech framework, which first discretizes the audio and text, and then relies on language models to demonstrate strong zero-shot speech synthesis capabilities. The training of TacoLM is also open-source to facilitate research in this direction. 
    \item We train the discrete speech language model leveraging on a moving average equipped gated attention mechanism (MEGA), which has a lightweight computation and storage compared to the vanilla multi-head attention.
    \item We design a novel gated cross-attention layer that efficiently links up information between text and the audio sequence, further improving the performance and efficiency.
\end{itemize}

%
%
\begin{figure}[t]
    \centering
    \includegraphics[width=1.0\linewidth]{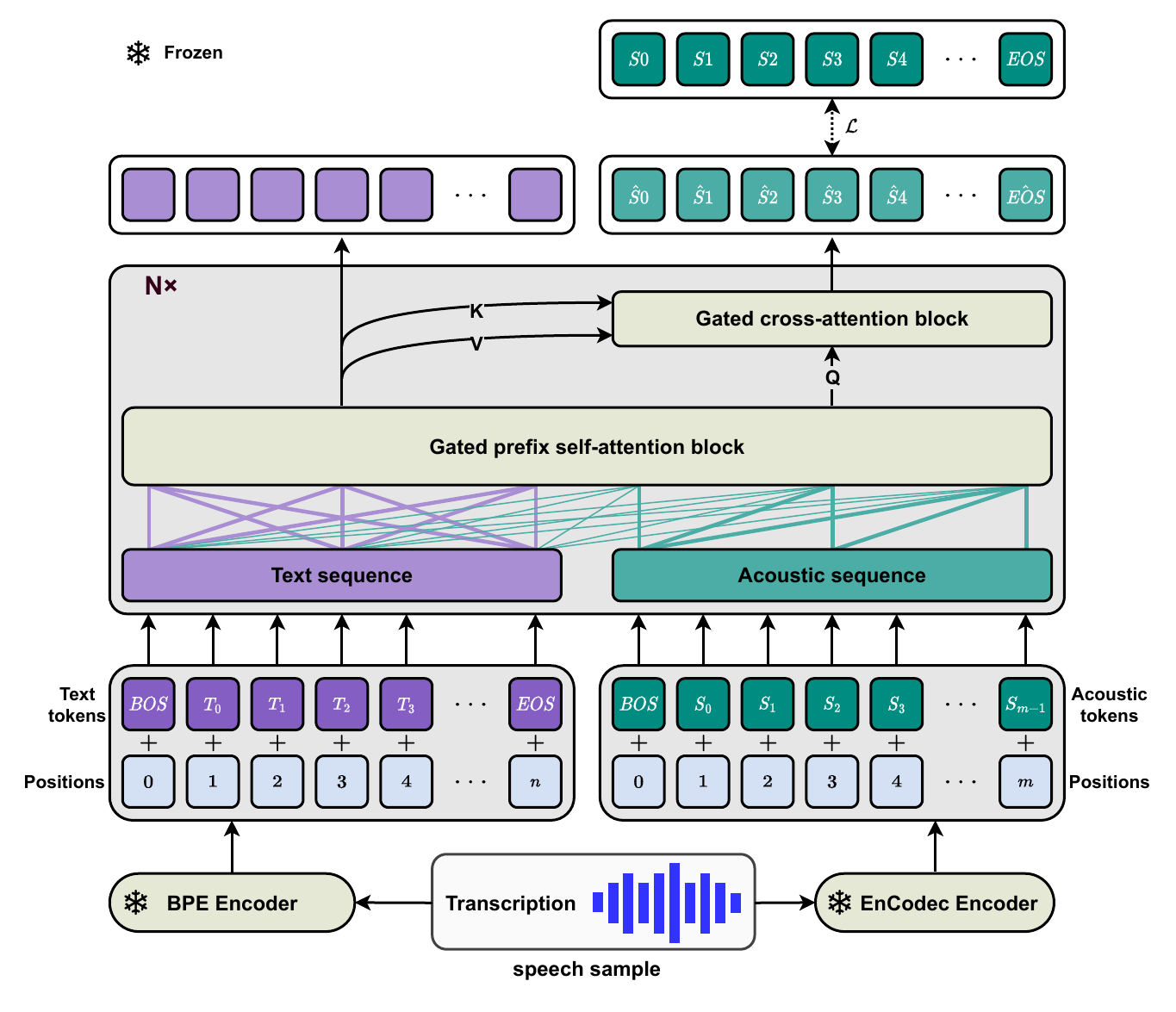}
    \vspace{-0.15in}
    \caption{\textbf{Framework Overview of the proposed TacoLM.} Gated prefix self-attention (GPSA) layers and gated cross-attention (GCA) layers are adopted in the AR model to generate the first layer of codec codes, while GCA layers are adopted in the NAR model to generate the rest layers of codes.  }
    \vspace{-0.12in}
    \label{fig:overall}
\end{figure}
%

\section{PROPOSED APPROACH}
\label{sec:pagestyle}

In this section, we introduce TacoLM, a TTS system with zero-shot speech synthesis capability. 
Similar to VALL-E, TacoLM consists of four main components: a text encoder, a neural audio codec, an autoregressive language model, and a non-autoregressive language model. In order to improve the training and inference efficiency and reduce the model size, we introduce two updates to the vanilla Transformer-based autoregressive language model, as depicted in Fig. \ref{fig:overall}. Firstly, 
a gated attention mechanism (namely MEGA~\cite{mega}) is used as an efficient drop-in replacement for regular multi-head attention. Moreover, we propose a gated cross-attention layer to improve the accuracy of synthetic speech from the paired text prompts, which further boosts the computation and storage efficiency of the autoregressive language model. 

\subsection{Model framework}
TacoLM operates in two stages. On the input side, we map the input text transcript and audio waveform to a sequence of semantic tokens, where a pre-trained neural audio codec model, EnCodec~\cite{encodec}, is used as the audio tokenizer. At the first stage, these tokens are input into the AR language model to generate the codec codes of the first quantizer of EnCodec. Subsequently, in the second stage, the NAR language model predicts the codes of the rest quantizers in parallel. In the inference process, given a target text sequence and a 3-second recording of an unseen speaker as a prompt, the neural codec decoder is able to synthesize high-quality waveform, keeping the acoustic environment and timbre of the speech prompt.
\newline

\noindent \textbf{Text encoder.} Text encoders are utilized to extract content representations from the text transcription. In this paper, byte pair encoding (BPE)~\cite{bpe} is used to extract discrete text representation. We use text data transcribed from the LibriSpeech 960h training set to directly train a BPE text encoder model, where the word vocabulary is set to 2000 and the Character coverage rate (CCR) is set to 1.0.
\newline

\noindent \textbf{Neural audio codec.} In this paper, we use the pre-trained neural audio codec model EnCodec~\cite{encodec} as the audio encoder for TacoLM. EnCodec is an efficient real-time neural audio compression model that generates high-fidelity audio samples. Through the RVQ module of EnCodec, speech tokens have a hierarchical structure: speech tokens from the low-level residual quantizers recover acoustic attributes, such as the speaker's identity and the coarse-grained content information, while the rest successive quantizers learn finer acoustic details. Each quantizer computes the residuals from the lower-layer quantizers. The EnCodec encoder converts the 24 kHz input waveform into 75 Hz discrete tokens, effectively reducing the sampling rate to $\nicefrac{1}{320}$. Each frame is represented using a residual vector quantization (RVQ) module with 8 hierarchical quantizers, each containing 1024 entries. In this setting, for each second of a 24 kHZ waveform, EnCodec's encoder synthesizes a matrix of $75\times 8$ entries as a discrete representation of the audio. 
\newline

\noindent \textbf{Autoregressive language model.} 
For discrete tokens from the first quantizer of the RVQ of EnCodec, we train an autoregressive language model. Its goal is to predict subsequent codewords from the first residual quantizer conditional on the target text sequence and the acoustic tokens. Formally, let $\mathcal{X}$ denote the transcribed text sequence, and $\mathcal{A}_{:,1}$ denotes the acoustic tokens of the first quantizer extracted
from the speech $\mathcal{S}$. The autoregressive prediction process of TacoLM can be formulated as:
\begin{align}
P \left( \mathcal{A}_{:,1} \,|\, \mathcal{X}; \theta_{AR} \right) = 
\prod_{t=1}^{T} \, p \left( \mathcal{A}_{t,1} \,|\, \mathcal{X}, \mathcal{A}_{<t,1}; \theta_{AR} \right)
\end{align}

\noindent \textbf{Non-autoregressive language model.} 
We obtain the codewords of the first quantizer through an autoregressive language model. In order to predict discrete tokens from the second to the last quantizer conditioning on the first layer, we train a non-autoregressive language model. Specifically, the prediction goal of the model can be expressed as:
\begin{align}
P \left( \mathcal{A}_{:,2:8} \,|\, \mathcal{X}; \theta_{NAR} \right) = 
\prod_{l=2}^{8} \, p \left( \mathcal{A}_{:,l} \,|\, \mathcal{X}, \mathcal{A}_{:,<l}; \theta_{NAR} \right)
\end{align}
The non-autoregressive language model has 8 independent acoustic embedding layers. The discrete speech tokens of the first $l-1$ layers are summed up to be used as inputs to the model to predict the tokens of the $l$-th quantizer.

%
\begin{figure}[t]
    \centering
    \includegraphics[width=0.85\linewidth]{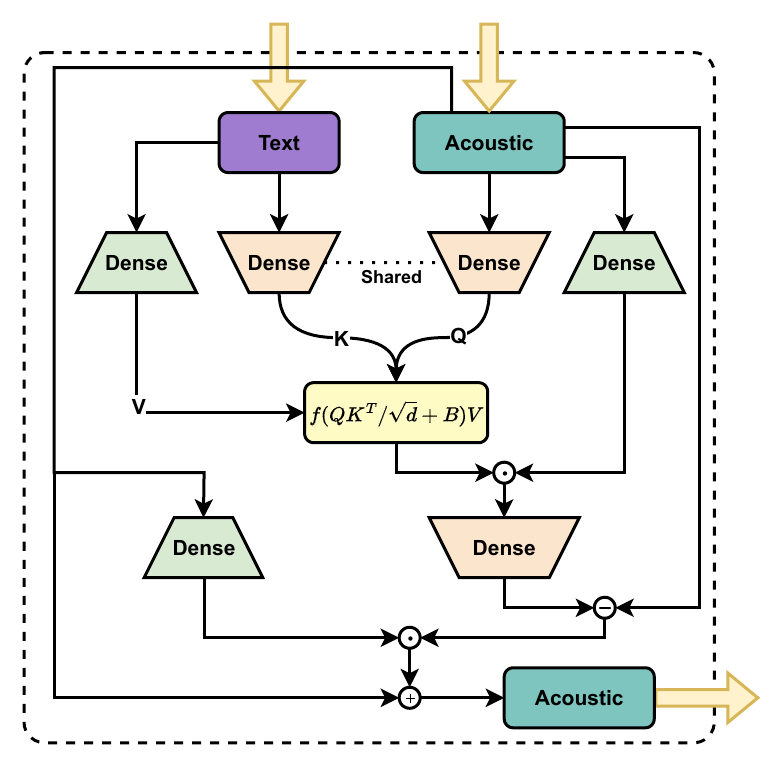}
    \vspace{-0.15in}
    \caption{
        \textbf{Illustration of the details of gated cross-attention layer.}
        $B$ in the yellow box refers to the relative position bias, where we use RoPE~\cite{rope} for position encoding. $d$ is the dimension of $Q$ and $K$.
    }
    \vspace{-0.12in}
    \label{fig:gca}
\end{figure}
%

\subsection{Gated Attention}
The key design in TacoLM is an autoregressive decoder-only network component based on a gated attention mechanism, which contains a gated prefix self-attention (GPSA) layer and a gated cross-attention (GCA) layer.

\noindent \textbf{Gated prefix self-attention layer.} Unlike previous works~\cite{audiolm,valle}, the decoder-only network of the TacoLM autoregressive language model first introduces the Moving Average Mechanism Equipped with Gated Attention (MEGA)~\cite{mega} as the drop-in replacement of traditional multi-head attention. The key idea of MEGA is to incorporate the classical exponential moving average (EMA) method into an attention mechanism for the whole sequence. With the benefit of single-head gated attention, MEGA has higher time and memory efficiency compared to multi-head self-attention. In this paper, we use bidirectional self-attention for text prefixes and unidirectional self-attention for discrete speech tokens to ensure the causality of audio generation.

\noindent \textbf{Gated cross-attention layer.} In the decoder-only causal language model, since unidirectional attention is applied to both the source sequence and the target sequence, less and less attention is focused on the source sequence as the length of the target sequence grows~\cite{fu2023decoder}. For the discrete speech language model, the attention degradation problem results in text mismatches when generating long sequences. In order to alleviate the problem, we propose a gated cross-attention layer after the GPSA layer, the structure of which is shown in Fig.\ref{fig:gca}. The gated cross-attention computes the key and value matrices of attention for text sequences and the query matrix of attention for acoustic sequences, which makes it focus on the text part and effectively mitigates the attention degradation problem since the key matrix is not affected by the growth of the target audio sequence.

\subsection{Inference}
In the inference process, we first encode the text transcription into a sequence of discrete codewords using BPE and the recorded audio prompts into an acoustic matrix using the EnCodec encoder. Both prompts are used for AR and NAR models. For the AR model, we employ the sample-based decoding method since beam search may cause the language model to enter an infinite loop, and greedy decoding is very unstable. In addition, the sample-based decoding method can significantly enhance the diversity of speech output. On the other hand, for the NAR model, we use greedy decoding to select tokens with the highest probability. Finally, we use the decoder of the neural codec to synthesize waveforms conditioned on a sequence of eight residual quantizer codewords.

\begin{table}[t]
    \small
    \centering
    \caption{
        \textbf{Objective and subjective evaluation for zero-shot TTS.} 
        For a fair comparison, we train VALL-E on the LibriSpeech dataset. $^\dagger$ indicates the audio is fed through EnCodec's encoder and decoder to eliminate the interference of the codec on the experimental results. The subjective evaluations are shown with a 95\% confidence interval (CI).
    }
    \resizebox{0.48\textwidth}{!}{
        \begin{tabular}{l|cccc}
            \toprule
            \textbf{Models}
             & SPK ($\uparrow$) & WER ($\downarrow$) & CMOS ($\uparrow$) & SMOS ($\uparrow$) \\
            
            \midrule
            \textbf{Ground Truth}$^\dagger$ & 0.9130 & 1.6211 & 0.30±0.04 & 4.50±0.05 \\
            
            \midrule[0.15pt]
            \textbf{VALL-E} & 0.8617 & 6.2461 & 0±0 & 3.50±0.11 \\
            
            \textbf{TacoLM} \textit{(ours)} & \textbf{0.8696} & \textbf{5.9560} & \textbf{0.12±0.08} & \textbf{3.75±0.11} \\
            
            \bottomrule
        \end{tabular}
    }
    \vspace{-0.125in}
    \label{tab:results}
\end{table}

\begin{table}[t]
    \small
    \centering
    \caption{
        \textbf{Model size, peak memory consumption, training speed, and inference latency comparison} of the autoregressive model with input length of 4K. All the inference experiments were conducted on a single NVIDIA GeForce RTX 3090 GPU.
    }
    \resizebox{0.48\textwidth}{!}{
        \begin{tabular}{l|cccc}
            \toprule
            \multirow{2}{*}{\textbf{Models}} &
                \multirow{2}{*}{\textbf{\#Param.}} &
                \multirow{2}{*}{\textbf{Mem.}} & 
                \textbf{Training Speed ($\uparrow$}) & \textbf{Inference Latency ($\downarrow$}) \\
             & & & (epochs/hour) & (RTF) \\
            
            \midrule
            \multirow{2}{*}{VALL-E} & 154.3M & 9.7GiB & 2.31 & 7.54 \\
             & $(1\times)$ & $(1\times)$ & $(1\times)$ & $(1\times)$ \\
             
            \midrule[0.25pt]
            \multirow{2}{*}{\textbf{TacoLM}} & \textbf{15.8M} & \textbf{3.0GiB} & \textbf{5.45} & \textbf{1.45} \\
             & $\mathbf{(0.10\times)}$ & 
             $\mathbf{(0.31\times)}$ & 
             $\mathbf{(2.37\times)}$ & 
             $\mathbf{(5.20\times)}$ \\
            \bottomrule
        \end{tabular}
    }
    \vspace{-0.125in}
    \label{tab:results2}
\end{table}

\section{EXPERIMENTS}

\subsection{Experimental Setup}

\begin{table*}[t]
    \small
    \centering
    \caption{
        \textbf{Detailed results of ablation study for TacoLM modules.}
    }
    \vspace{-0.1in}
        \begin{tabular}{c|ccccccc}
            \toprule
            \multirow{2}{*}{\textbf{Models}} &
                \multirow{2}{*}{\textbf{SPK}($\uparrow$)} &
                \multirow{2}{*}{\textbf{WER}($\downarrow$)} &
                \multirow{2}{*}{\textbf{PPL}($\downarrow$)} &
                \multirow{2}{*}{\textbf{\#Param.}} &
                \multirow{2}{*}{\textbf{Mem.}} & 
                \textbf{Training Speed}($\uparrow$)  & \textbf{Inference Latency}($\downarrow$)  \\
             & & & & & & (epochs/hour) & (RTF) \\
            
            \midrule
            \multirow{2}{*}{\textbf{TacoLM}} & 
                \multirow{2}{*}{\textbf{0.8696}} & 
                \multirow{2}{*}{\textbf{5.9560}} & 
                \multirow{2}{*}{\textbf{13.76}} & 
                \textbf{15.8M} & \textbf{3.0GiB} & \textbf{5.45} & \textbf{1.45} \\
             & & & & $\mathbf{(0.10\times)}$ & 
             $\mathbf{(0.31\times)}$ & 
             $\mathbf{(2.37\times)}$ & 
             $\mathbf{(5.20\times)}$ \\
             
            \midrule[0.25pt]
            TacoLM & 
                \multirow{2}{*}{0.8632} & 
                \multirow{2}{*}{6.5157} & 
                \multirow{2}{*}{13.85} & 
                18.5M & 4.4GiB & 3.16 & 2.36 \\
             w/o GCA & & & & $(0.12\times)$ & $(0.46\times)$ & $(1.37\times)$ & $(3.19\times)$ \\

            \midrule[0.25pt]
            TacoLM & 
                \multirow{2}{*}{0.8617} & 
                \multirow{2}{*}{6.2461} & 
                \multirow{2}{*}{13.95} & 
                154.3M & 9.7GiB & 2.31 & 7.54 \\
            w/o GCA\&GPSA & & & & $(1\times)$ & $(1\times)$ & $(1\times)$ & $(1\times)$ \\
            
            \bottomrule
        \end{tabular}
    \vspace{-0.125in}
    \label{tab:ablation}
\end{table*}

\subsubsection{Data}
We train TacoLM on the multi-speaker English speech dataset LibriSpeech~\cite{librispeech}. Specifically, we use ~\texttt{train-clean-100}, ~\texttt{train-clean-360}, and ~\texttt{train-other-500} as the training set, which contains a total of 960 hours of 16kHz English speech. For the evaluation, we select only the test samples from ~\texttt{test-clean} and ~\texttt{test-other} with a duration between 4 and 10 seconds following VALL-E~\cite{valle}. 

\subsubsection{Implementation Details}
To extract acoustic tokens, we use the official open-source EnCodec~\cite{encodec} checkpoint, which is trained on a variety of 24 kHz monophonic audio data. Since LibriSpeech is a 16 kHZ speech dataset, we first upsample the speech data to 24kHZ to feed EnCodec. For text tokens, We use SentencePiece~\cite{sentencepiece} as the text tokenizer to extract the byte pair encoding from the transcription of the speech corpus. 
In the AR model, we interleave 6 GPSA layers and 6 GCA layers. More specifically, the AR model has 6 blocks, each of which contains a GPSA layer and a GCA layer. The NAR model is similar to AR, except it uses a bidirectional attention mask and stacks 12 GPSA layers. For both AR and NAR, we set embedding dimension to 384, hidden state dimension to 384, feed-forward layer dimension to 768, damped EMA dimension to 24, key and value projection dimension to 240, and a dropout of 0.1. The activation function used is silu (sigmoid linear unit) \cite{silu}. All models are trained in parallel using 8 NVIDIA GeForce RTX 3090 GPUs with a batch size of 8192 tokens per GPU, learning a total of 240k steps. We use the AdamW optimizer with $\beta_1=0.9$, $\beta_2=0.999$, $\epsilon=10^{-9}$. For the learning rate schedule, we linearly increase the learning rate from zero to a peak of $10^{-3}$ for the first 12k updates, followed by a linear decay. The weight decay is 0.05 and clip-norm is 1.0.

\subsubsection{Baselines}
We compare zero-shot speech synthesis performance with the recent state-of-the-art zero-shot TTS system VALL-E, which was trained on the 60k hours Librilight dataset~\cite{librilight} in the original paper. We reproduce VALL-E\footnote{https://github.com/Ereboas/TacoLM/tree/valle} and re-train it on the LibriSpeech 960h training set. For a fair comparison, we process the ground truth speech by feeding through EnCodec's encoder and decoder, to eliminate the interference of the audio codec. For each test sample, we provide the first 3 seconds of speech as the speech prompt and ask the model to synthesize the speech of the specified transcription.

\subsubsection{Evaluation Metrics}
We use speaker similarity (SPK) and word error rate (WER) as the objective metrics, and conduct the mean opinion score (MOS) evaluation as the subjective metrics. Specifically, for SPK evaluation, we use the state-of-the-art speaker verification model, WavLM-TDNN~\cite{wavlm}, to assess speaker similarity between the speech prompt and the synthesized speech. WavLM-TDNN predicts similarity scores in the range of [-1,1], with larger values indicating higher similarity between the synthesized speech and the speech prompt. When the similarity is greater than 0.86, WavLM-TDNN claims that two speech samples come from the same speaker. We also evaluated the accuracy of the synthesized speech to the text prompt. We perform Automatic Speech Recognition (ASR) on the generated speech and calculate the WER relative to the original transcription. In this work, we use the current advanced model Conformer-Transducer~\cite{conformer} as the ASR model. For the subjective evaluation, we selected 60 testing pairs randomly, each listened to by a minimum of 15 listeners. The listeners are then asked to rate the audio samples based on naturalness or speaker similarity. We analyze the comparative mean option score (CMOS) in terms of naturalness, and the similarity mean option score (SMOS) which measures speaker similarity. SMOS is rated on a scale of 1 to 5 on a 0.5-point increment, where higher is better. CMOS ranges from -1 to 1, and higher scores indicate better system performance compared to the baseline.

For our main concern of efficiency issue, we recorded the number of model parameters, training memory usage, training speed, and inference latency of the models. The inference latency is measured using the real-time factor (RTF) metric, which means how much time the model needs to generate one second of audio. The inference length is fixed at 4500 tokens, which support EnCodec with a target bandwidth of 24kbps to synthesize 15 seconds of speech.

\subsection{Result}
We present the zero-shot TTS evaluation results in Table~\ref{tab:results}. TacoLM outperforms the baseline with statistical significance (p-value $<$ 0.05 from the Wilcoxon signed-rank test). From Table~\ref{tab:results2}, TacoLM has only 0.10 times the model size of VALL-E and 0.31 times the training memory, and achieves 2.37 and 5.20 times the training and inference speeds, respectively. Furthermore, due to the use of Rotational Position Embedding (RoPE)~\cite{rope} for training, TacoLM can also be naturally extrapolated to any longer sequence than the training sequences during inference. To summarize, TacoLM demonstrates very competitive performance in terms of both effectiveness and efficiency.

\vspace{-0.04in}
\subsection{Abaltion Study}
In this section, we conduct ablation experiments to evaluate the effectiveness of the modules in TacoLM. Specifically, we analyze two variants of the model: TacoLM with GCA layers replaced by GPSA layers (TacoLM w/o GCA), and TacoLM with both GCA and GPSA layers replaced by vanilla multi-head attention layers (TacoLM w/o GCA\&GPSA). For all three settings, we use a total of 12 attention layers with the same hyper-parameters as TacoLM or VALL-E. In addition, we compared the perplexity (ppl) of the three AR language models. The results are presented in Table~\ref{tab:ablation}. We can observe that the inclusion of the GPSA layer leads to a significant reduction in the time and space costs, while causing little or no degradation in performance. On the other hand, when replacing the GCA layers, we observe a degradation in the speaker similarity (SPK) and speech accuracy (WER) performance, which indicates that cross-attention plays a crucial role in enhancing the accuracy of synthesized speech with respect to the source text prompt.

\vspace{-0.04in}
\section{Conclusion \& Limitation}
\vspace{-0.02in}
In this paper, we proposed TacoLM, a zero-shot text-to-speech approach towards computational efficiency and synthesis accuracy, built upon the foundation of the AR + NAR two-stage neural codec language modeling.
We introduced the gated prefix self-attention layer to reduce the memory and storage requirements while improving training and inference speeds. We further proposed a novel gated cross-attention layer to increase the accuracy of synthesized speech with respect to the source text. 
Our experimental results demonstrated that TacoLM not only outperformed the existing state-of-the-art zero-shot TTS system across a range of evaluations, but also significantly reduced memory and storage usage, and enhanced computational efficiency.
This study still has some limitations. TacoLM relies on two-stage modeling, limiting its convenience for end-to-end training. In addition, limited to resources, TacoLM has not yet been trained and tested on a larger corpus. This prevents us from verifying its scalability, and is left for future works.

\section{Acknowledgements}
This work was supported by the National Natural Science Foundation of China  (No. 62206171 and No. U23B2018), Shanghai Municipal Science and Technology Major Project under Grant 2021SHZDZX0102, the International Cooperation Project of PCL and Alibaba Innovative Research Program.

\bibliographystyle{IEEEtran}
\bibliography{mybib}

\end{document}